\documentclass[%
 reprint,
 amsmath,amssymb,
 aps,
pra,
]{revtex4-2}
\usepackage{graphicx}
\usepackage{dcolumn}
\usepackage{bm}
\usepackage{booktabs}
\usepackage{siunitx}
\usepackage[textsize=tiny, textwidth=15mm]{todonotes}

\usepackage{upgreek}
\usepackage[english]{babel}
\usepackage{silence}
\usepackage{subfig}
\usepackage{float}
\usepackage{wrapfig}
\usepackage{caption} 

\usepackage{comment}
\usepackage{amsmath}
\begin{document}

\preprint{APS/123-QED}

\title{Role of the Local Electric-Field in High-Field Conditioning of DC Electrodes: Numerical and Experimental Insights}

\author{Victoria M. Bjelland}
 \email{victoria.bjelland@cern.ch}
 \altaffiliation[Also at ]{NTNU}
\author{William L. Millar}%
 \author{Walter Wuensch}
 \email{walter.wuensch@cern.ch}
\affiliation{Espl. des Particules 1, 1217 Genève, CERN
}%

\author{Morten Kildemo}
\affiliation{Department of Physics, NTNU, Heogskoleringen 5
}%

\date{\today}

\begin{abstract}
Conditioning, the progressive increase of voltage-holding through the controlled application of fields, is an important and widely used process for bringing high-field and high-voltage devices up to their full operating parameters. Here, a study is presented on how conditioning can vary within a device, specifically, when there is a spatial variation in the surface electric field. What has been observed in high-field pulsed direct-current electrodes and radio-frequency structures is that locations exposed to higher fields exhibit a greater tendency to breakdown, but this increase is counteracted by an increased conditioning rate. This interplay explains the observed breakdown locations and provides important insights into the mechanisms underlying both conditioning and breakdown. This study combines Monte Carlo simulations with experimental results from pairs of high-field electrodes with a radially varying surface electric field. Results are presented from a high-field pulsed DC system, in which the position of each breakdown during conditioning was recorded by triangulation using a pair of cameras, and the results are compared with Monte Carlo simulations.
\end{abstract}

\maketitle

\section{Introduction}\label{introduction}
In many high-field and high-voltage devices, it is common practice to gradually increase the operating voltage while monitoring for arcs, rather than immediately commencing operation at the design voltage. Doing so typically increases the likelihood of establishing high electric fields without arcing. This procedure, known as conditioning, is an essential step in the commissioning process for vacuum interrupters and normal-conducting particle accelerator cavities and waveguide components ~\cite{vi_conditioning,clara_condition,xband_comps}.

The investigation of field-driven conditioning has both practical and theoretical benefits. Conditioning procedures are time-intensive processes and typically carry an inherent risk of component damage. An improved understanding of the process can aid in identifying material, manufacturing, and treatment choices that mitigate the need for conditioning, reduce associated risks, and facilitate optimization of existing procedures. A prominent example is the high-gradient radio-frequency (RF) cavities developed for applications ranging from cancer treatment to future collider concepts, i.e. flash therapy and CLIC~\cite{app13085021, Aicheler:1500095, CLICReport}. Cavities of this kind operate at surface fields up to $\sim$220~MV/m, and generally require substantial conditioning periods~\cite{Aicheler:1500095, CLICReport,  MARTINEZREVIRIEGO2025103164, NuriaConditioning}.

To shed light on the mechanisms that underlie conditioning, a Monte Carlo simulation was previously developed at CERN~\cite{millar:linac2022-mopori24}. The report presented here describes the design of two electrodes based on simulations performed with this tool, followed by a dedicated experiment conducted in the CERN pulsed DC system~\cite{PROFATILOVA2020163079,wuensch2017}. To investigate the field dependence of conditioning, a tapered electrode design was employed to establish an inhomogeneous surface electric field. The electrodes were conditioned to a peak surface electric field of approximately 80~MV/m, and the breakdown positions were recorded on a pulse-to-pulse basis. The spatial variation in the surface electric field, combined with precise temporal and spatial breakdown data, serves two main purposes. Firstly, the results shed light on the mechanisms underlying conditioning and breakdown by clearly illustrating the interplay between the applied field that produces a conditioning effect and the field that drives breakdown, and by providing information on the relative strength of each phenomenon. Secondly, the results serve as an important dataset against which models of conditioning and breakdown, such as the one presented herein, may be benchmarked.

\section{Features of High-Field Conditioning}\label{conditioning generally}

To date, numerous experiments have been conducted to explore the mechanisms associated with high-field vacuum arcs, field holding, and conditioning. Parameters that have been studied include the device geometry, material, preparation techniques, operating frequency (including DC), the gaps between electrode surfaces, and the high-field pulse length and repetition rate~\cite{braun2003, Profatilova2019, dolgashev2010, Wuensch2023, Lucas_reprate, psi_paper}. Recently, DC and RF tests have also been performed at cryogenic temperatures, and this approach now forms the basis of a proposed future collider concept~\cite{marek_2020, marek_2025, nasr_2021, Schneider_2024, c3} . Unlike normal-conducting structures, where RF conditioning is commonly employed to reduce breakdown rates within the accelerating structure itself, superconducting cavities rely primarily on meticulous surface preparation and contamination control to suppress breakdown-inducing mechanisms. RF conditioning is therefore generally applied only to the input couplers as part of its acceptance test~\cite{hli:srf17-mopb020,Pierini2023SRF}. In the context of existing facilities, attempts have been made to discern the effects of radiation exposure on high-field performance~\cite{serafim2024,serafim2025}.

A comprehensive set of models that describe the different experimental signatures, dependencies, and stages of breakdown has been developed. The breakdown process first involves material dislocation dynamics driven by the stress from the applied surface electric field, followed by the formation of surface features, their evolution driven by surface dynamics and finally heating, electron emission and field driven neutral atom emission and finally plasma formation. A review of the current status of the understanding of the breakdown process is given in ~\cite{Wuensch:2025muq}. An improved understanding of conditioning will contribute to further progress in understanding breakdown.

\subsection{Conditioning Procedures}
For the purposes of this work, conditioning is defined as the increased propensity to establish high fields without arcing or increases in internal pressure that would prohibit operation in normal-conducting vacuum devices. In the early stages of operation, some high-field devices, particularly RF cavities, are commonly limited by multipacting and elevated pressure levels rather than by arcing alone~\cite {Mitra2012, Papadopoulos2017, Xiao_2019, Piquet2022}. In one case, an algorithm was developed specifically to condition cavities using the pressure level as the process variable~\cite{Woolley2015Thesis}. However, after this period, the rate at which the operating field can be increased is often limited primarily by the occurrence of vacuum arcs. Historically, conditioning procedures were often carried out on anecdotal or empirical evidence rather than a precise scientific understanding of the process. However, automated algorithmic approaches to conditioning are now commonplace, and offer a more consistent and systematic approach to component testing and commissioning~\cite{Woolley2015Thesis, clara_condition,Benedetti2022,Trachanas2023,tom2025}.

Generally, conditioning procedures, both manual and algorithmic, are based on the breakdown rate, i.e., the number of vacuum arcs that have occurred within a chosen temporal window, or equivalently, the number of pulses. The pulse width and field level are then adjusted to maintain the target breakdown rate until the final operating parameters are reached. Practically speaking, this means progressively increasing the operating field while monitoring for arcs to maintain a permissible rate of arcing. However, additional action is also occasionally required to respond dynamically in the event of, e.g., an elevated arc rate or otherwise anomalous behavior.

\subsection{Characteristics of Conditioning}

\begin{figure}[!htbp]%
    \centering
    \includegraphics[width=1\columnwidth]{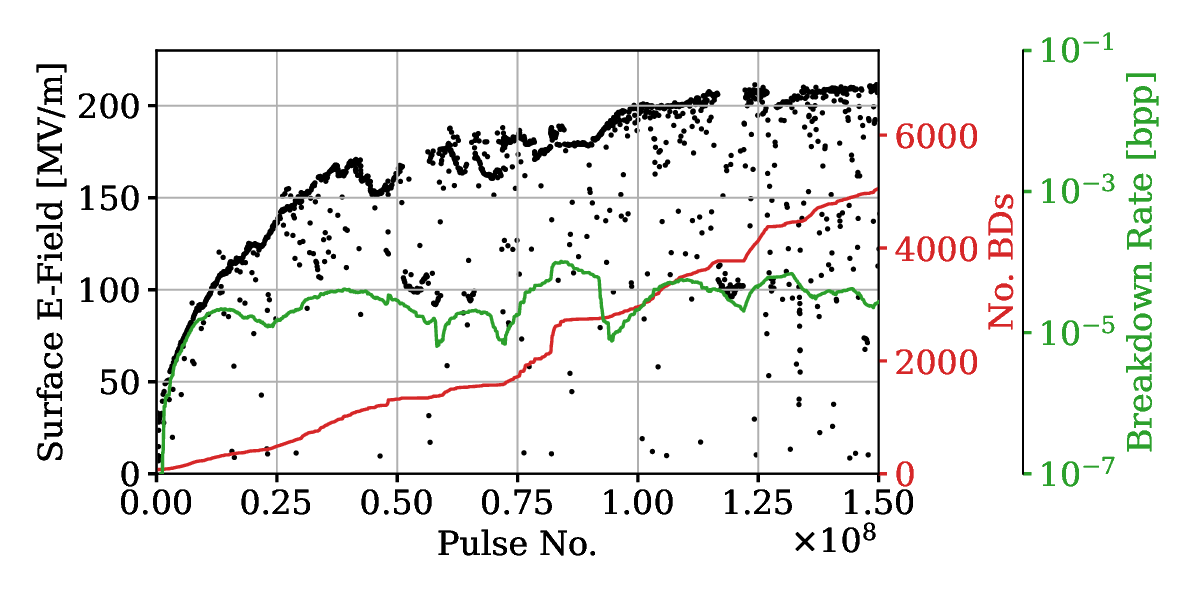}%
    \caption{\label{subfig:electrode_conditioning}Conditioning curves of a \SI{12}{\giga\hertz} copper RF cavity (T24 PSI N1~\cite{psi_paper}) that was tested in CERN's high-gradient test facility (a) and a copper DC electrode that was tested in the LES with a gap size of \SI{60}{\micro\meter} (b), illustrating a typical asymptotic increase in applied field. Note that the x-axes differ in scale by an order of magnitude. }
    \label{fig:typical_conditioning}
\end{figure}

During conditioning procedures, if a constant breakdown rate is maintained, the rate at which the field can be increased typically decreases, leading to asymptotic behavior. To illustrate this behavior, the conditioning of an X-band accelerating structure and a DC electrode are shown in Fig~\ref{fig:typical_conditioning}.
The approach of progressive field ramping has been shown to reliably enable operation at high fields with a reduced probability of arcing, and a major theoretical objective of this work is to gain insight into the mechanism underlying this behavior.


Long-term tests performed on X-band ($\sim$\SI{12}{\giga\hertz}) RF cavities have shown that the conditioning of similar devices proceeds most comparably in terms of cumulative high-field pulses rather than the cumulative number of breakdowns~\cite{degiovanni_conditioning}. This result has led to the suggestion that the conditioning effect, i.e., the increased field-holding capability, is produced by the application of high fields and is not solely a consequence of the arcs. Therefore, DC electrodes are conditioned in a staircase fashion, where the applied electric field is increased more gently. The conditioning approach is explained in more detail in Section~\ref{exp setup}. This result forms the basis for much of the work conducted in this manuscript and is discussed further in Section~\ref{sim and des elect}.

\subsection{Characteristics of Vacuum Arcs}

As they constitute one of the main limitations on stable high-field operation, significant effort has gone into improving the understanding of vacuum arcs, as well as predicting and mitigating their occurrence~\cite{obermair2022}. In various RF and DC high-field tests, it has been shown that if an arc occurs following a quiescent period, it is common for one or several more arcs to occur in quick succession, and that these groups of events are likely to occur in close proximity~\cite{wuensch2017}. This observation has led to the grouping of arcs into primary or trigger events, which occur stochastically, and secondary events, which are thought to be consequences of the former~\cite {wuensch2017, cahill2018, psi_paper}.

\begin{figure}[H]
    \centering
    \includegraphics[width=\linewidth]{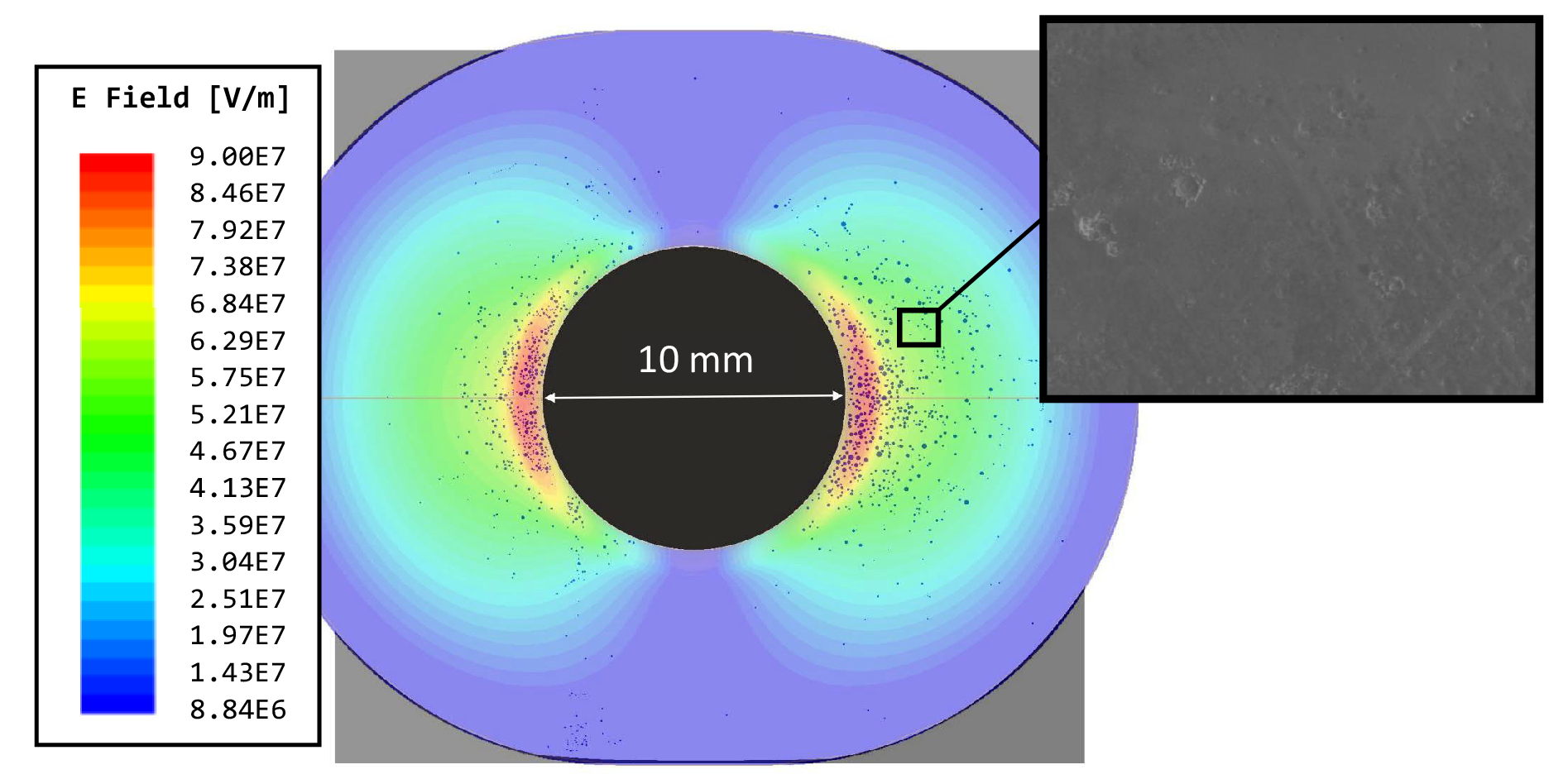}
    \caption{Electric field distribution on the second cell of the CLIC crab cavity with the breakdown locations superimposed (shown by the dark blue spots). Figure taken from a previous publication~\cite{millar:linac2022-mopori24}. The test procedure and post-test analysis are reported in full elsewhere~\cite{crab_2020}.}
    \label{fig:crab}
\end{figure}

To date, several quantities have been studied as predictors of device performance. Examples include the group velocity in RF devices~\cite{Adolphsen2001, Adolphsen2005}, the ratio of the power flow to the iris size in accelerator cavities~\cite{Wuensch2006}, and the so-called modified Poynting vector, $S_c$~\cite{sc_paper}. Although RF cavity studies have shown that, for a given breakdown rate, similar peak magnetic fields are obtained for different designs~\cite{dolgashev2010}, post-test optical examinations have shown that the arcs predominantly occur where the surface electric field is highest~\cite{crab_2020}. An example is shown in Fig.~\ref{fig:crab}, where the surface of an X-band RF cavity that operates in the TM$_{110}$ mode is presented with the surface electric field distribution and the arc locations overlaid. In this mode, the regions where $S_c$, the surface electric field, and the surface magnetic field are well separated, and a post-test optical examination showed that the majority of breakdown sites were present where the electric field is highest~\cite{crab_2020}.

More broadly, a comprehensive analysis of long-term

\noindent RF cavity tests found the breakdown rate to scale with the electric field $E$, the thirtieth power, or $E^{30}$~\cite{sc_paper}. More recent tests have shown similar power law relations, with reported exponents ranging from 28 to 35~\cite{psi_paper, pablo_2025}. However, this scaling emerges globally, and weaker dependencies have been reported across regions of inhomogeneous field within a given device~\cite{crab_2020}, as shown in Fig.~\ref{fig:crab}. In pulsed RF devices, the breakdown rate is commonly quoted as scaling with the pulse length, $\tau$, to the fifth power, $\tau^5$~\cite{sc_paper,degiovanni_conditioning}. However, recent attempts to measure this dependency have yielded lower exponents, and the interpretation of such measurements is occasionally complicated by the use of non-rectangular pulse shapes~\cite{matsumoto2011, Wu_2021, psi_paper}. In a previous study, the potential influence of the order in which the measurements are conducted was also noted~\cite{psi_paper}.

\subsection{Motivation}

Generally, arcs serve as the feedback mechanism during conditioning procedures, with the occurrence of an arc indicating that the field level is sufficiently high to produce a non-negligible conditioning effect. During such procedures, the field can gradually be increased while maintaining a constant rate of arcing. Conversely, if the applied field is held constant, measurements have shown that the arc rate gradually decreases. These observations suggest that the application of high electric fields increases the likelihood of arcing and that the arc rate and/or the conditioned state of the surface continuously evolve over the course of a given measurement. The objective of this work is to investigate the interplay between these two effects and to provide information on the relative strength of each mechanism.

In multicell traveling wave (TW) RF cavity tests, localization of arc sites to a specific cell is often accomplished in situ by employing a time-of-propagation-based method~\cite{Rajamaki2016}. However, localization to a specific region on the surface is generally not possible without a post-test visual examination. Consequently, the exact spatial and temporal evolution of the arc distribution on the surface is lost. In some conditioning studies using DC electrodes, precise real-time breakdown localization is possible, though only samples that support homogeneous electric fields have been tested thus far~\cite {Profatilova2020}. In this study, this limitation is addressed by testing a pair of electrodes that establishes an inhomogeneous surface electric field while recording the arc locations and the order in which they occur. The electrodes were designed with the aid of a Monte Carlo model of conditioning~\cite{millar:linac2022-mopori24}. The objective of the optimization process is to select a geometry that, after conditioning, would exhibit a gradual decrease in arc density across its surface.

Therefore, to the authors' knowledge, this dataset represents the first time that the positions of arcs and the order in which they were accrued have been recorded simultaneously in a pulsed device operating with an inhomogeneous surface electric field. The results also serve to validate the model used to design the electrodes, thereby confirming its suitability for more pragmatic applications, such as optimizing existing conditioning procedures.

Notably, at the High Energy Accelerator Research Organization (KEK), a camera was integrated into a high-power cavity test, allowing direct in-situ observation of breakdowns as they occurred~\cite{abe2016}. However, this arrangement was operated in a continuous wave regime and focused on mechanisms associated with breakdown triggers, rather than on the evolution of the arc distribution and the conditioned state of the surface, which are the focus of this work. 

\begin{figure}[H]
    \centering
    \includegraphics[width=\linewidth]{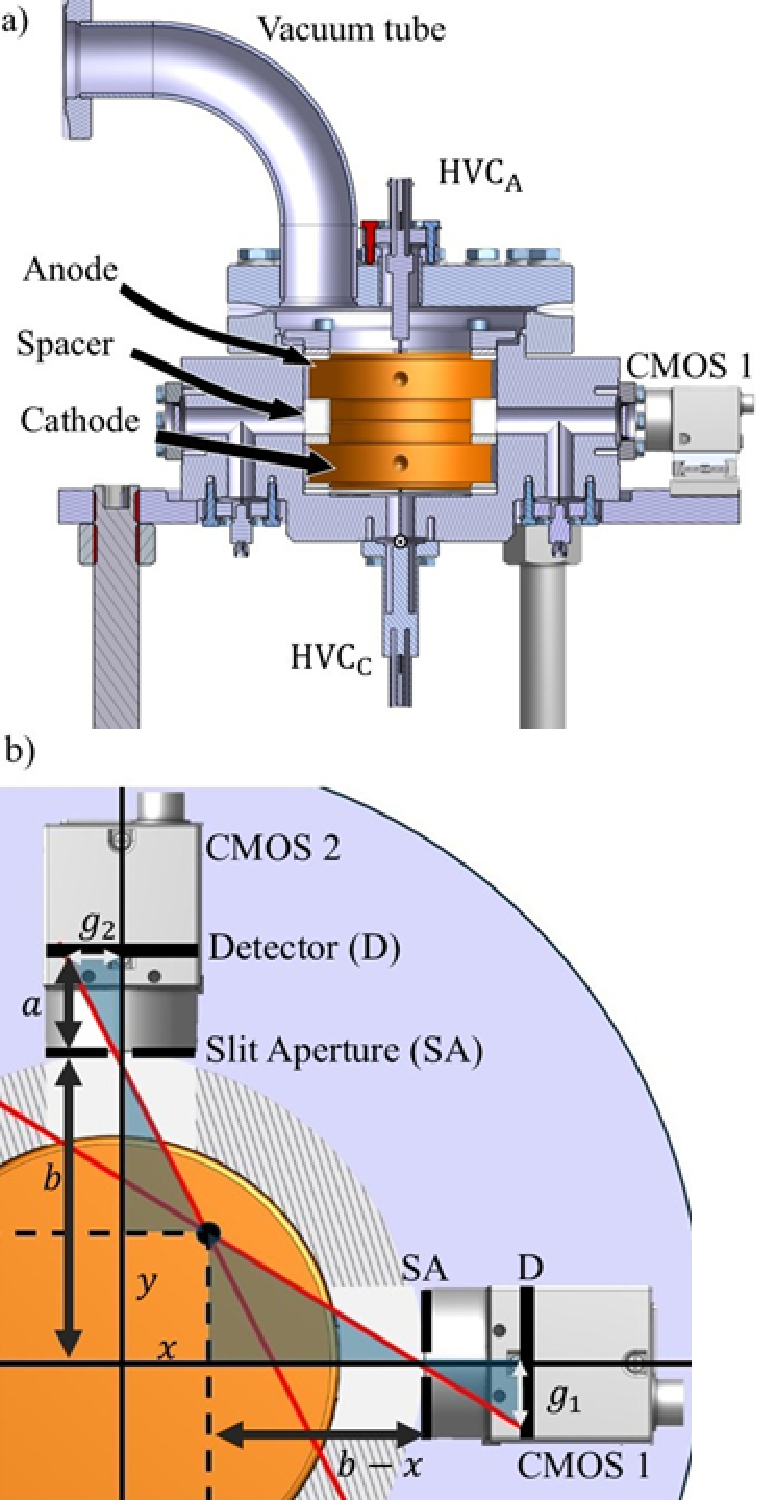}
    \caption{Cross-sectional views of the LES shown from the side (a) and from above (b). a) The electrodes are separated by an insulating spacer, creating a micrometer precision gap. Electrical contact to the anode and cathode is made via the HVC$_A$ and HVC$_C$, respectively. Two CMOS cameras are attached to the four symmetrical viewports on the system. Only one is shown here, as the other is hidden behind the system. b) Breakdown localization system. When a breakdown occurs, it creates a flash of light originating from $(x,y)$. Two red lines can be drawn, going through the vacuum breakdown and the slit apertures (SA), and hitting the detectors (D). This creates two similar triangles, marked in blue, from which the original breakdown location can be calculated using the known system constants $a$ and $b$ and the measured quantities $g_1$ and $g_2$. The equations are given elsewhere~\cite{Profatilova2019}. }
    \label{fig:ExperimentalSetup}
\end{figure}

\section{Experimental Setup, Conditioning Procedure and Spatial Light Triangulation}\label{exp setup}

The experimental measurements were performed in a vacuum chamber fitted with a series of interchangeable electrodes. Due to the large size of the electrodes, this configuration is commonly referred to as the Large Electrode System (LES)~\cite{PROFATILOVA2020163079, Saressalo_2020_1, Saressalo_2020_2,9203211, Engelberg:2020epz}. Figure~\ref{fig:ExperimentalSetup}~a) shows a cross-sectional view of the setup. Inside, the two electrodes are separated using a precision-machined insulating spacer. The spacer rests on the cathode, while the anode sits atop the spacer. Two pin-type high-voltage connector (HVC) feedthroughs, specialized for delivering and receiving power to and from the interior of a vacuum chamber. HVC$_\text{A}$ and HVC$_\text{C}$ connect to the anode and cathode, respectively. On the side, one of the two CMOS cameras is shown. The second is positioned perpendicular, at the back of the chamber.  

The chamber is evacuated using a single turbomolecular pump to achieve vacuum levels down to $10^{-9}$~mbar. Pressure is monitored along the vacuum line using a cold-cathode vacuum gauge and recorded at a sampling rate of 3~Hz.

The electrodes typically comprise a flat machined ($R_a < 20$~nm) metallic cathode and anode to ensure a spatially uniform surface electric field.  Depending on the chosen spacer, the typical electrode gaps are 20, 40, 60, or 100~$\upmu$m. Here, the 60~$\upmu$ spacer was chosen because it is a common gap size. The gap distance is verified by capacitance measurements, as the electrodes act as a parallel-plate capacitor. The exposed radii of the cathode and anode are 30~mm and 20~mm, respectively. These dimensions remove edge-related field enhancement, which would localize breakdown events at the electrode periphery due to the increased surface electric field~\cite{Saressalo_2020_1}. The electrodes consist of oxygen-free (OFE) Cu electrodes, which were heat-treated in a hydrogen atmosphere up to 800~$^\circ$C to extend the conditioning time~\cite{Profatilova:2703709}. 

To achieve a non-uniform electric field, the anode was machined to have a flat-topped conical shape, a fustrum. The overall gap was determined by the spacer used. This leads to the gap height $d(r)$ that is a function of the radius
\begin{equation}
    d(r) = 
    \begin{cases}
    60~\upmu\text{m}, & r\leq 6.5~mm \\
    60~\upmu\text{m}+(r-6.5~\text{mm})\tan(\theta), & r > 6.5~mm
    \end{cases}
\end{equation}
where $\theta = \tan^{-1}({(60-h_2)~\upmu\text{m}}/{23.5 \text{mm}})$. The gap height at the anode edge, $h_2$, was determined by simulations and is covered in detail in Section~\ref{sim and des elect}. Consequently, the surface electric field can be found as 
\begin{equation}\label{eq:applied voltage}
E_{\text{surface}}(r) = \frac{V_{\text{applied}}}{d(r)},
\end{equation}
ensuring a uniform surface electric field when $r \leq 6.5$~mm and a radially decaying when $r > 6.5$~mm.

During testing, short high-voltage pulses are applied to HVC$_\text{A}$ provided by a pulsed high-voltage generator (EPULSUS-FPM1-10)~\cite{electronics13010101}, while HVC$_\text{C}$ is grounded. The generator employs a Marx circuit configuration, charging ten capacitors in parallel before discharging them in series, hence the name Marx generator. The LES Marx generator can deliver pulses up to 10~kV and 50~A, with pulse widths ranging from 500~ns to 100~ms. The generator is powered by an external supply.

When a vacuum breakdown occurs during a pulse, the chamber transitions from an insulating vacuum between the electrodes to a low-impedance state, causing a rapid surge in current demand of up to 50~A within 10~ns. The breakdown is detected by the sudden current demand from the Marx generator, which prompts the applied pulse to stop after 600~ns. During this, the pressure will spike to the order of 10$^{-6}$~mbar before decaying exponentially back to stable values. Therefore, the system will wait 3~s before pulsing continues to let the vacuum recover. During recovery, the applied voltage is reduced to 20~$\%$ of its pre-breakdown value and then logarithmically ramped back up to the conditioning setpoint~\cite{Peacock:2023euy}. If the breakdown rate (BDR), the number of breakdowns per pulse, exceeds a preset threshold (typically $10^{-5}$~bds/pulse), the system actively lowers the applied voltage until the BDR is met~\cite{Peacock:2023euy}. A collection of conditioning studies using this experimental setup can be found in the literature~\cite{Peacock:2023euy, Profatilova:2703709, PhysRevAccelBeams.23.033102, serafim2024, serafim2025}.

The vacuum breakdown also produces brief flashes of light that resemble a point source and persist for only a few microseconds~\cite{Kovermann:2010vea}. For copper, the light emission in the LES occurs as a broadband signal across the visible spectrum, with its peak at 750~nm~\cite{Peacock:2023euy}. Two perpendicular cameras, each equipped with a 100~$\upmu$m wide slit aperture, are gathering the optical data from the chamber, as seen in Fig.~\ref{fig:ExperimentalSetup}~b), operating continuously and synchronized with an exposure time of $\tau = 3$~s. 

The slit apertures restrict the transmitted light so that only the central portion of the optical beam reaches the detectors, producing a Gaussian intensity profile on each camera. A line drawn through the center of the Gaussian profile and the corresponding slit aperture defines the optical line of sight, which also passes through the breakdown location. By having two cameras, the breakdown location ($x,y$) can be found as the point at which the two lines cross, as shown in Fig.~\ref{fig:ExperimentalSetup}~b). The two measurable quantities, $g_1$ and $g_2$, along with the system constants, $a$ and $b$, are used to find the origin location with equations given in~\cite{PROFATILOVA2020163079}.

\section{\label{sim and des elect} Simulation and Design of a Tapered Electrode}

Previously at CERN, a Monte Carlo-based model was developed to simulate the conditioning and operation of high-field devices on a pulse-to-pulse basis~\cite{millar:linac2022-mopori24}. The model is deliberately constructed to be theory-agnostic and relies on two main assumptions:

\begin{enumerate}
    \item The application of an electric field produces a local conditioning effect, i.e., exposure to high electric fields increases the surface's propensity for establishing electric fields without arcing.
    \item The probability of arcing scales strongly with the local electric field, with higher fields corresponding to an increased probability of arcing.
\end{enumerate}

In addition to more pragmatic motivations, such as optimizing conditioning procedures, one of the key objectives was to investigate the interplay between these two counteracting effects and the discrepancies in empirical scaling laws that have been reported. Although the model, and several preliminary studies conducted with it, have been discussed elsewhere~\cite{millar:linac2022-mopori24}, it has since undergone several modifications and has not been applied to DC devices. As it was employed in the design of the electrodes reported in this paper, and therefore forms the basis for much of the work reported herein, an overview is provided. Several peculiarities specific to the simulation of copper electrodes operating in a pulsed DC regime are also discussed.

\subsection{Overview of the Model}
\label{sec:model_overview}
The surface of the high-field device to be simulated is first meshed, yielding an array of $n$ elements. Taking $i$ as the element number, each element has the following attributes:

\begin{itemize}
\item $a_{i}$ - The ratio of the mesh element area to the total simulated area. 
\item $k_{i}$ - The ratio of the local electric field to the peak electric field for a given operating voltage or power level. 
\end{itemize}

\noindent With this approach, the high-field conditioning of any arbitrary geometry and surface electric field distribution is permitted. Additionally, the use of an inhomogeneous mesh permits the use of larger elements in regions where little activity, e.g., infrequent arcs or little conditioning, is expected to reduce the computational cost of the simulation. 

Numerous experiments worldwide have shown that the voltage or power level required to establish a given breakdown increases asymptotically during conditioning procedures. Based on this observation, the model assumes that for a given pulse length and reference breakdown rate, $P_{Ref}$, there is a maximum, material-dependent surface electric field, $E_L$, in MV/m that a device can be conditioned to operate at. More succinctly, this assumption forbids the unphysical procedure of conditioning to arbitrarily high field levels or low breakdown rates. 

To capture the conditioning effect, each surface element is also a conditioning state, $E_{S,i}$, which refers to the field level in MV/m at which the mesh element is conditioned to operate. Each element is also assigned a state enhancement factor, $\psi_i$, which is initially selected randomly from an appropriately tailored Gaussian distribution, and reselected each time the element accrues an arc. Depending on the applied peak electric field, $E_O$, which also has units of MV/m, the conditioning state of the surface elements is increased on a pulse-to-pulse basis as

\begin{equation}
\label{eq:conditioningRate}
\Delta E_{S,i} = \gamma_i \cdot  \frac{E_{O}\cdot k_{i}}{E_{S,i}} \cdot \left[1 - \frac{E_{S,i}}{E_{L,i}} \right],
\end{equation}

\noindent where $\gamma_i$ is a constant with units of V/m that allows the model to be fit to experimental data, and varies depending on the material, operating frequency, and pulse length. Generally, most breakdowns that occur during conditioning procedures occur in the preliminary stages, as the applied field is increased to the target level~\cite{degiovanni_conditioning, psi_paper}. As the pulse length is typically fixed during this preliminary stage, the use of a fixed value for $\gamma_i$ is sufficient for the studies reported in this work. 

Although some variation is present, in well-conditioned copper RF devices the breakdown rate has typically been observed to increase with the applied electric field to the thirtieth power, i.e., $BDR \propto E_O^{30}$, ~\cite{sc_paper, psi_paper, pablo_2025}, and investigating whether this scaling accurately reproduces the behavior of other devices, particularly those constructed from different materials or with different operating frequencies, is another objective of the model. Based on this empirically derived scaling, the probability of a particular grid element arcing on a given pulse, $P_{BD, i}$, is expressed as

\begin{equation}
\label{eq:breakdownProbability}
P_{BD,i} =  a_i \cdot \left[\frac{E_{O}\cdot k_{i}}{E_{S,i}\cdot \psi_i}\right]^{30} \cdot P_G, 
\end{equation}

\noindent where $P_G$ is a probability that is applied to the grid elements individually to obtain a global breakdown rate that is comparable to $P_{Ref}$ in the later stages of conditioning. To derive a suitable value for $P_G$, the probability of a pulse not resulting in an arc is more convenient to define. Assuming the system is fully conditioned everywhere, and all $\psi_i$ values are equal to one, 

\begin{equation}
\label{eq:p_grid_calculation1}
1 - P_{Ref} = \prod_{i=1}^n (1 - P_G \cdot a_i \cdot k_i^{30}).
\end{equation}

Although a solution can be obtained using iterative numerical methods, a reasonable approximation can also be obtained. Taking the natural logarithm of both sides of Eq.~\ref{eq:p_grid_calculation1} yields

\begin{equation}
\label{eq:p_grid_calculation2}
\ln(1 - P_{Ref})  = \sum_{i=1}^n \ln(1 - P_G \cdot a_i \cdot k_i^{30}) .
\end{equation}

However, $P_{Ref} \ll 1$, with typical values being less than $1 \times 10^{-4}$. Therefore, by taking the first term of the Taylor series, a reasonable first-order approximation may be obtained as

\begin{equation}
\label{eq:p_grid_calculation3}
\ln(1 - P_{Ref}) \approx - P_{Ref}.
\end{equation}

\noindent As $P_G \cdot a_i \cdot k_i^{30} < P_{Ref}$ for any scenario with more than one grid element, the same holds true for the right side of Eq.~\ref{eq:p_grid_calculation2}. Rearranging the terms then yields

\begin{equation}
\label{eq:p_grid_calculation4}
P_G \approx  \frac{P_{Ref}}{G},
\end{equation}

\noindent where G is a geometry factor, which corresponds to the surface integral of the surface electric field to the thirtieth power, defined

\begin{equation}
\label{eq:G_factor}
G =  \sum_{i=1}^n  a_i \cdot k_i^{30}.
\end{equation}

Consequently, Eq.~\ref{eq:breakdownProbability} dictates that if all $\psi_i$ values are equal to one, every surface element has been conditioned to its limit, $E_{L,i}$, and the device is operated at this level, the breakdown rate will be approximately equal to $P_{Ref}$. This ensures that the performance in simulation is consistent with the experimental data from which $E_{L,i}$ and $P_{Ref}$ were selected, regardless of the mesh resolution. In low-field regions, the assumption that the surface has been fully conditioned is not necessarily correct. However, the probability of breakdown scales extremely strongly with the applied field, as per Eq.~\ref{eq:breakdownProbability}, and so the error that this effect introduces is small.

In several previous studies, an alternative approximation was used~\cite{millar:linac2022-mopori24}, which instead relied on the assumption that

\begin{equation}
\label{eq:alternative_pgrid1}
1 - P_{Ref} \approx (1-P_G)^{G}.
\end{equation}

\noindent Rearranging this expression, $P_G$ can then be defined as

\begin{equation}
\label{eq:alternative_pgrid2}
P_G = 1 - \left( 1 - P_{Ref} \right)^{\displaystyle \left(\frac{1}{ G}\right)}
\end{equation}

This approximation emerges from taking the natural logarithm of the left-hand side in Eq.~\ref{eq:alternative_pgrid1}, exponentiating the result, and, as $P_g \ll 1$, once again taking the first term of the Taylor expansion to show

\begin{equation} \label{eq:alternative_pgrid3}
\begin{split}
(1 - P_G)^{G} &= \exp\left(\ln(1 - P_G) \cdot G \right)\\
 & \approx \exp{\left( -P_G \cdot{G}\right)}\\
 \therefore 1 - P_{Ref}  & \approx \exp{\left( -P_G \cdot{G}\right)}, 
\end{split}
\end{equation}

\noindent and therefore, taking the natural log once again and employing the approximation given in Eq.~\ref{eq:p_grid_calculation3}, the same approximation as Eq.~\ref{eq:p_grid_calculation3} emerges as 

\begin{equation}
\label{eq:alternative_pgrid4}
- P_{Ref} \approx  -P_G \cdot G.
\end{equation}

The definition in Eq.~\ref{eq:alternative_pgrid2} holds if the variation in $a_i \cdot k_i^{30}$ products is small, and therefore is suitable for devices where the mesh and electric field distribution are uniform, or in cases where regions of lower electric field are compensated by much larger mesh cells. However, the form given in Eq.~\ref{eq:p_grid_calculation4} yields an improved result when larger cell-to-cell variations are present, and is therefore more broadly applicable. For this reason, the Eq.~\ref{eq:p_grid_calculation4} definition was employed for the studies presented in this work. 

Regarding the selection of $P_{Ref}$ and $E_L$, to accurately reproduce a device's behavior, the chosen values should ideally be taken from late-stage measurements conducted with the pulse length used in the early stages of conditioning. Measurements at multiple pulse lengths are often conducted during more comprehensive high-field tests~\cite{psi_paper, xiaowei_2017}, and so this aspect of the model is generally not problematic. Additionally, as $\gamma_i$ is presumed to be constant for a given set of device parameters, only a single appropriate test is required to derive an appropriate value. 

Additionally, the $P_{Ref}$ and $E_L$ values are somewhat interchangeable, provided the $\gamma$ value has been appropriately selected so as to provide the correct dynamic behavior. Hence, to tune the model, the device's approximate final performance must be known a posteriori. Once the material properties are known and verified, however, the model may be applied to other geometries and devices by simply rescaling $P_{Ref}$ to account for changes in surface area and electric-field distribution in the new design. Taking a trivial example, if two identical reference devices were to be simulated in parallel, the surface area, and therefore the reference breakdown rate, should also double for a given field level. More generally, the chosen reference breakdown rate is adjusted to provide a scaled reference breakdown rate, $P_{Ref}^{S}$, as

 \begin{equation}
\label{eq:ScaledPref}
P_{Ref}^{S} = \frac{A^S \cdot G^S}{A \cdot G} \cdot P_{Ref},
\end{equation}

\noindent where $A^S$ and $A$ are the surface area of the device being simulated, and the device from which $P_{Ref}$ was taken, respectively. Similarly, $G^S$ and $G$ are the geometric factors, as per Eq.~\ref{eq:G_factor}, of the device being simulated and the reference device, respectively. The probability applied to the grid elements of the new device, $P_G^S$, is then defined like in Eq.~\ref{eq:p_grid_calculation4}, as

\begin{equation}
\label{eq:ScaledPg}
P_G^S \approx  \frac{P_{Ref}^S}{G^S}.
\end{equation}

\indent and hence, for the new device Eq.~\ref{eq:breakdownProbability} becomes

 \begin{equation}
\label{eq:ScaledbreakdownProbability}
P_{BD,i} =  a_i \cdot \left[\frac{E_{O}\cdot k_{i}}{E_{S,i}\cdot \psi_i}\right]^{30}  \cdot P_{G}^{S}.
\end{equation}

\subsection{Electrode Design}\label{Electrode Design}

To provide a reasonable foundation for the design of an appropriate electrode, the model described in Section~\ref{sec:model_overview} was first tuned by eye to approximately reproduce a standard, flat electrode \textbf{(Mention gap size)} based on the results of previous high-field copper electrode tests conducted at CERN. A simulation of the conditioning of a copper electrode using the field ramp shown in Fig.~\ref{subfig:electrode_conditioning} is shown in Fig.~\ref{fig:tuned_electrode_simulation}, and chosen model settings are summarised in Table~\ref{tab:les_tuning_parameters}. The total number of breakdowns accrued in the simulation is lower than in the experimental case, as the anomalous cluster of breakdowns that occurred around 750 million pulses in Fig.~\ref{subfig:electrode_conditioning} and necessitated a reduction in operating field during the experiment was not reproduced by the model.

\begin{figure}[htbp!]
    \centering
    \includegraphics[width=\linewidth]{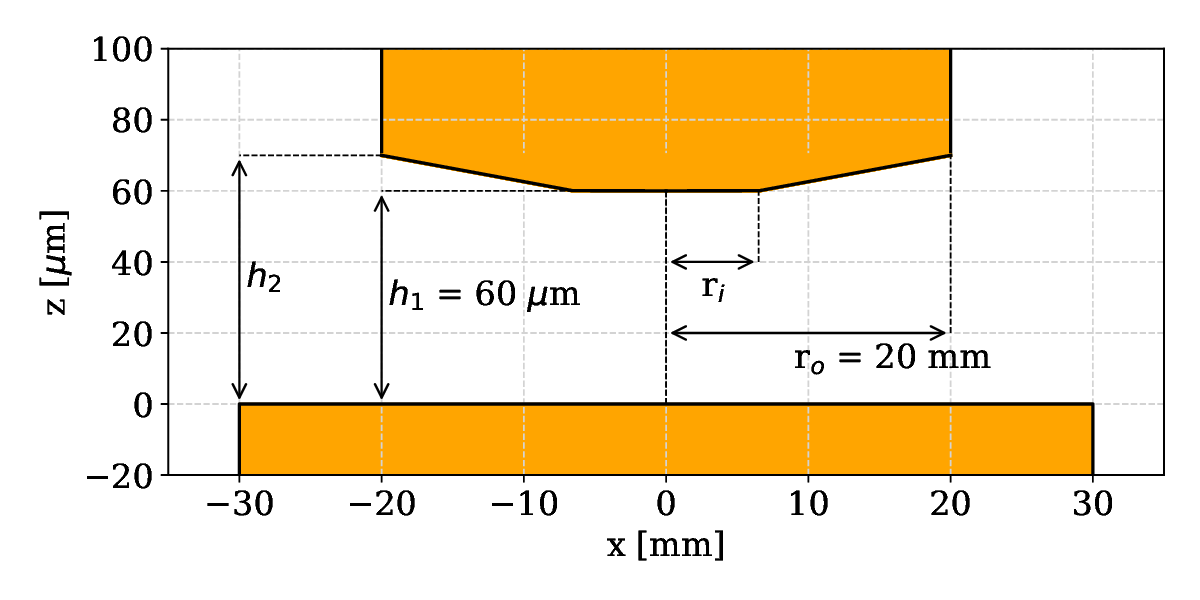}
    \caption{Cross-section of the tapered electrode geometry. Due to the small gap size, units of $\upmu$m are used on the z-axis, while units of mm are used on the x-axis. Dimensions $h_1$ and $r_0$ are shown for completion, but were fixed at \SI{60}{\micro\meter} and \SI{20}{\milli\meter} respectively and not varied in simulation.}
    \label{fig:electrode_cross_section}
\end{figure}

A cross-section of the frustum electrode, including the relevant dimensions, is shown in Fig.~\ref{fig:electrode_cross_section}. The LES system and existing spacers are designed to accommodate anodes with a radius $r_0$ of \SI{20}{\milli\meter} and, as described in Section~\ref{exp setup}, a gap size $h_1$ of \SI{60}{\micro\meter} is a standard and commonly used configuration for electrode tests.

Simulations were then performed using different combinations of $r_i$ and $h_2$, while $r_0$ and $h_1$ were held constant. The objectives were to identify a combination of values that produce a clearly discernible decrease in the number of arcs per unit area in the outward radial direction and to ensure that the flat central region remains sufficiently large to serve as a reference for normalization. Given the electrode's azimuthal uniformity, the surface was meshed using concentric bands. The electric field associated with each concentric band was calculated by evaluating the electrode separation at the band's radial midpoint, thereby yielding a band-averaged field. An illustrative example using a relatively coarse mesh is shown in Figure~\ref{fig:electrode_mesh}. 

In practice, some deviation from this idealized approximation arises due to field enhancement at electrode edges and the presence of fringe fields, particularly at the outermost band. However, arc-induced damage during high-field testing occurs predominantly on the cathode surface, and so the impact of these effects is mitigated by machining the relevant features on the anode rather than the cathode. 

\begin{table}[!htpb]
\caption{The chosen dimensions for the frustum electrode experiment, following the naming scheme shown in Figure~\ref{fig:electrode_cross_section}. The slope angle is 0.0422~°, and results in a field reduction 24~$\%$ at the edge of the active area on the cathode.} \label{tab:electrode_dimensions}
\centering
\begin{tabular}{c|cc}
\toprule\toprule
Quantity & Value & Unit \\
\midrule
$h_1$   & 60 & $\upmu$m \\
$h_2$   & 70 & $\upmu$m  \\
$r_{i}$ & 6.5 & mm  \\
$r_{o}$ & 20 & mm  \\
\bottomrule
\end{tabular}
\end{table}

\begin{figure}[H]
    \centering
    \includegraphics[width=\linewidth]{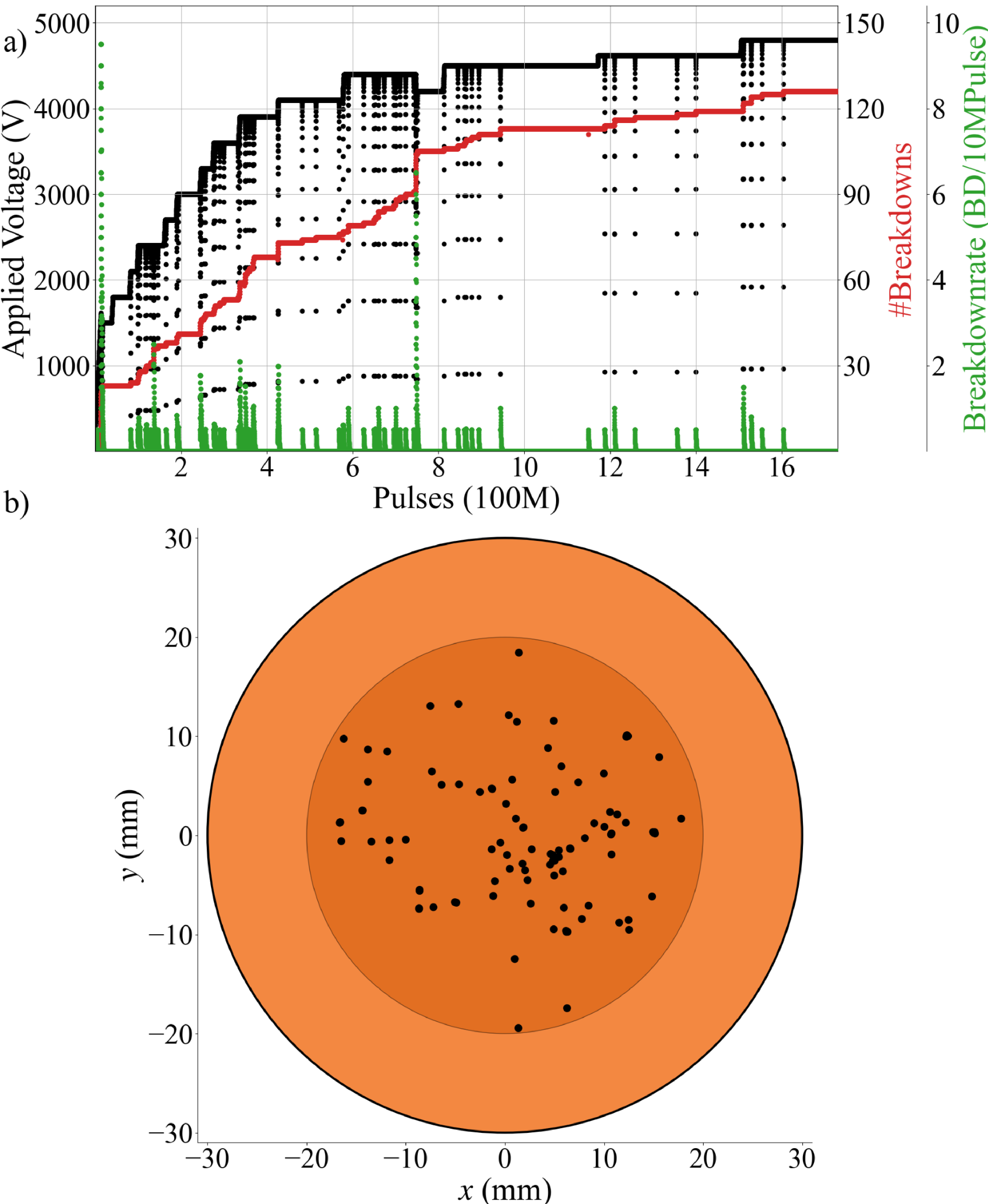}
    \caption{(a) Conditioning curve for the sloped electrode pair. The black line is the applied voltage, the red line is the number of accumulated breakdowns, and the green line is the breakdown rate. The center of the electrode reached $E_{center} = 80$~MV/m after 20~M pulses. (b) Accumulated breakdown locations over the cathode surface. The darker circle on the surface represents the effective field-exposed area, which is equivalent to the anode surface with a radius of 20~mm.} 
    \label{fig2:Conditioning}
\end{figure}

\section{Experimental Results}
\label{exp results}
\subsection{Conditioning and spatial breakdown locations}

As briefly introduced in Section~\ref{exp setup}, short high voltage pulses of 1~$\upmu$s at 1~kHz are applied on the electrodes. To condition them, the start voltage was set to 300~V (5~MV/m at the center of the electrode) and rose to 1500~V (25~MV/m at the center of the electrode) as the conditioning goal, in 10~V increments per 100 000 pulses. After letting the system run flat to ensure a lower breakdown rate, the conditioning goal was increased by 300~V, and the process was repeated until reaching a final applied voltage of 4800~V, as seen in Fig.~\ref{fig2:Conditioning}~a). The black line represents the applied voltage across the electrodes, $V_{\text{applied}}$, the red line shows the cumulative number of breakdowns, and the green line corresponds to the BDR.

The surface electric field is calculated from Eq.~\ref{eq:applied voltage}. When the center of the electrode reaches a surface electric field of $E_{\text{center}}$, the outer edge experiences a reduced field strength, given by $E_{\text{edge}} = 0.86 E_{\text{center}}$. Since the center was conditioned to a surface electric field of 80~MV/m, the outermost edge was conditioned to 68.8~MV/m. The spatially located breakdown locations over the cathode surface are shown in Fig.~\ref{fig2:Conditioning}~b), where the darker region is the field-exposed area equivalent to the area covered by the anode.

\begin{figure*}
    \centering
    \includegraphics[width=\linewidth]{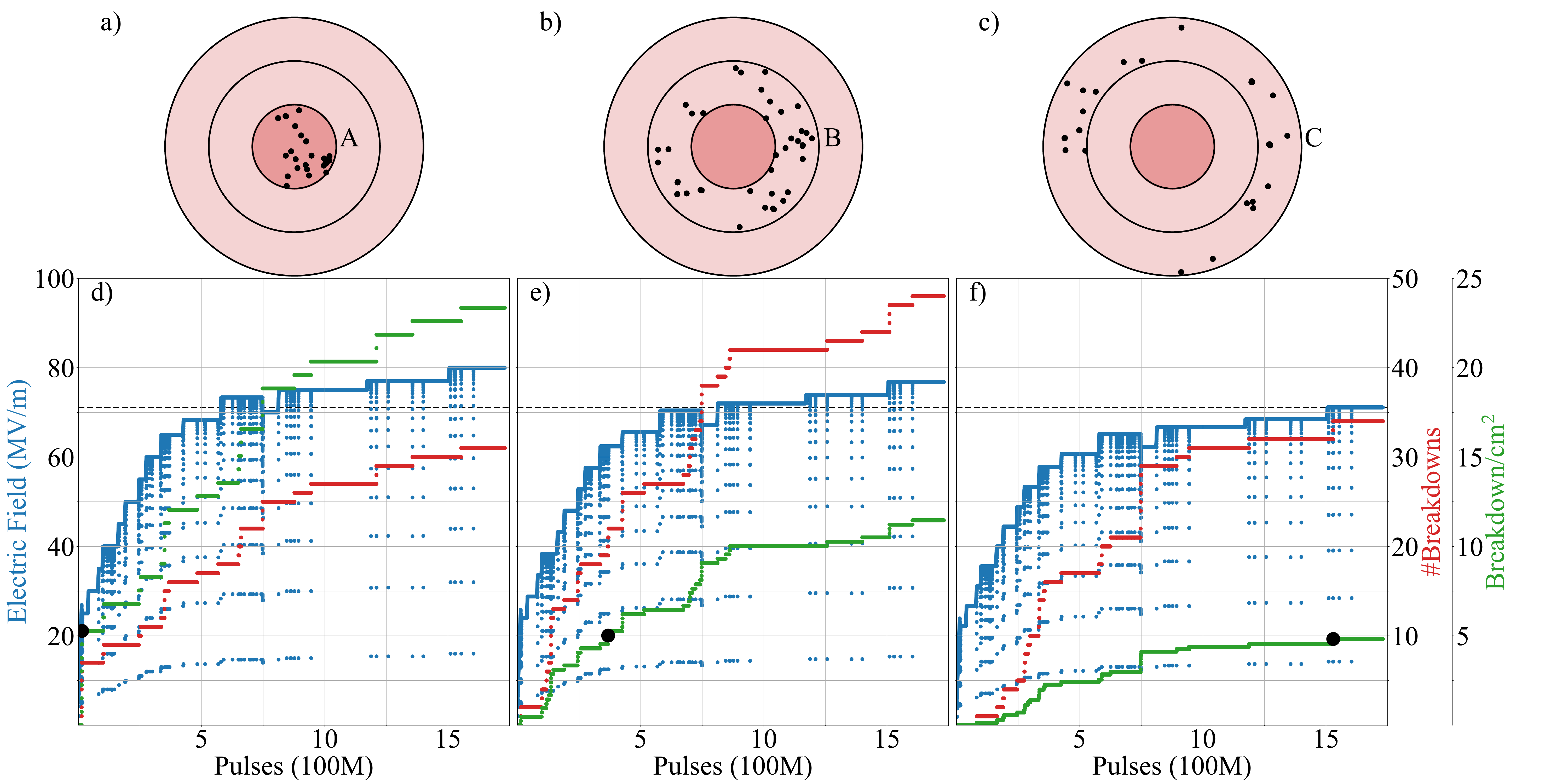}
    \caption{a-c) Spatially located breakdown positions, divided into sections A (inner), B (middle), and C (outer). d-f) Spatially divided conditioning where the blue lines are the average electric field of each region, the red lines are the accumulated breakdowns in each region, and the green lines are the corresponding breakdown density of each region.}
    \label{fig4:Division3}
\end{figure*}

\subsection{Spatial Division of Conditioning}

To better understand the relationship between conditioning and breakdown behavior, the electrode surface was divided into three sections: A, B, and C. Spatially resolved breakdown events are categorized according to these sections, as shown in Figs.~\ref{fig4:Division3}~a–c). Section A (inner) corresponds to the flat region of the anode, which exhibits a uniform surface electric field and covers $r\leq$6.5~mm. Section B (middle) covers the area from the edge of section A to the midpoint of the slope, given as 6.5~mm$<r\leq$13.25~mm. Section C (outer) covers the last part of the exposed electric-field area for $r>13.25$~mm. Unlike the simulated example, the experimental electrode had fewer breakdowns. Therefore, it was divided into 3 main areas instead of 10.

\begin{figure}[t]
    \centering
    \includegraphics[width=\linewidth]{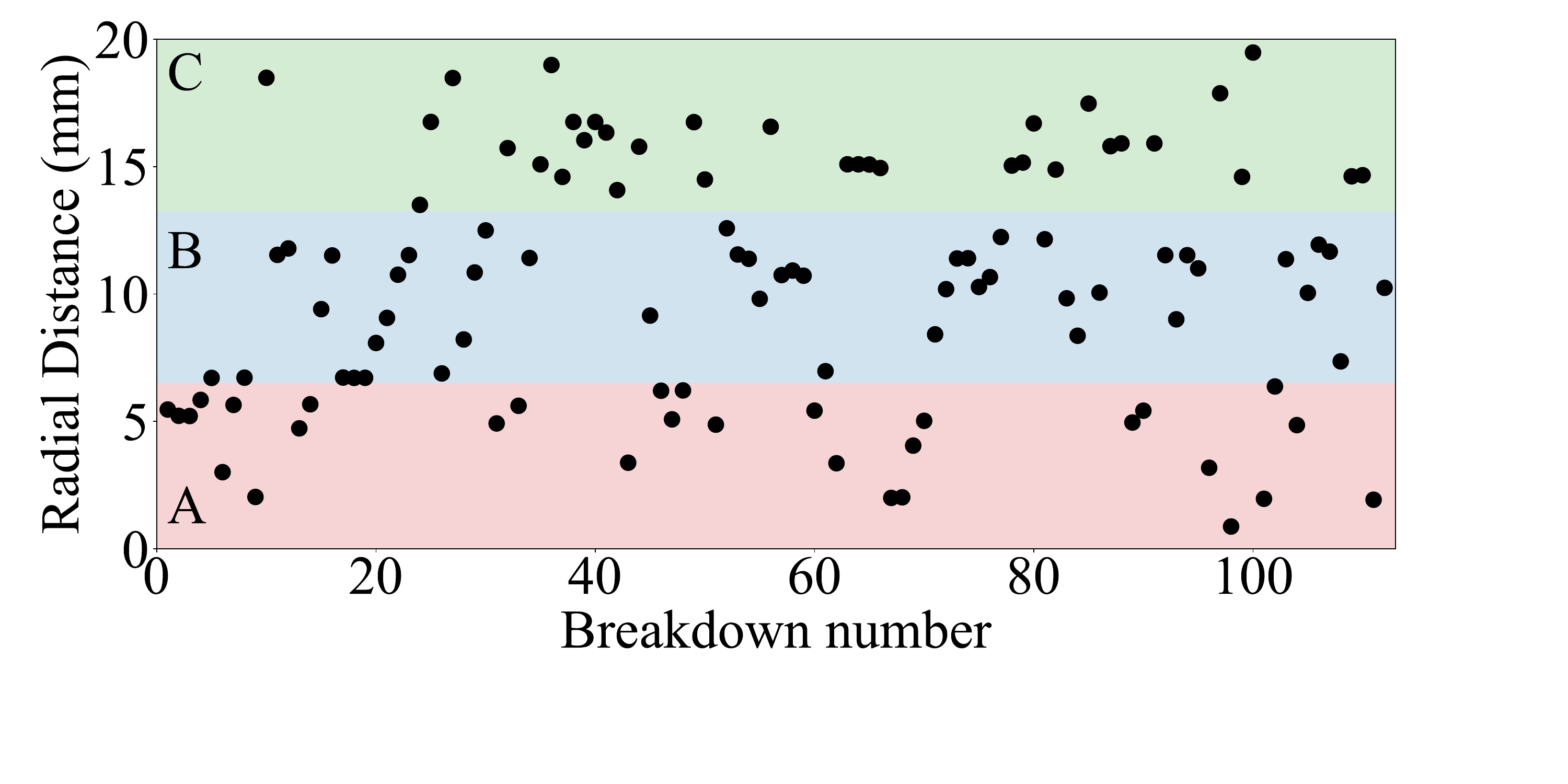}
    \caption{Breakdown number versus the distance from the center of the electrode to the breakdown. Regions A, B, and C are shown in red, blue, and green, respectively.}
    \label{fig:placementovertime}
\end{figure}

Figures~\ref{fig4:Division3}~d–f) present the conditioning data segmented by the spatial regions. In each figure, the blue line represents the spatially averaged surface electric field, the red line shows the cumulative number of breakdowns, and the green line indicates the spatial breakdown density (BDD), which is the number of breakdowns divided by the section's surface area. The maximum (average) surface electric fields achieved in Sections A, B, and C were $E_{\text{inner}} = 80~\text{MV/m}$, $E_{\text{middle}} = 77~\text{MV/m}$, and $E_{\text{outer}} = 71~\text{MV/m}$, respectively.

The number of accumulated breakdowns per section is highest in region B, followed by region C and then region A, due to the increasing surface area as the radius increases. In fact, the corresponding final breakdown densities were $24.0~\text{BD/cm}^2$, $12.2~\text{BD/cm}^2$, and $5.0~\text{BD/cm}^2$ for regions A, B, and C, respectively.

Figure~\ref{fig:placementovertime} shows the radial distance as a function of the breakdown number. The earliest breakdowns were predominantly concentrated in section A, followed shortly by events in section B and then in section C. After approximately 30 breakdowns, the spatial distribution becomes more uniform, with breakdowns occurring across the entire electrode surface, with no spatial correlation observed. This trend is also evident in Figure~\ref{fig4:Division3}, where the breakdown number in each region increases throughout the conditioning process. However, due to differences in area, the increase in BDD is weaker at greater radial distances.

\subsection{Simulated results against experimental observations}

\begin{figure}[tbh!]
    \centering
    \includegraphics[width=\linewidth]{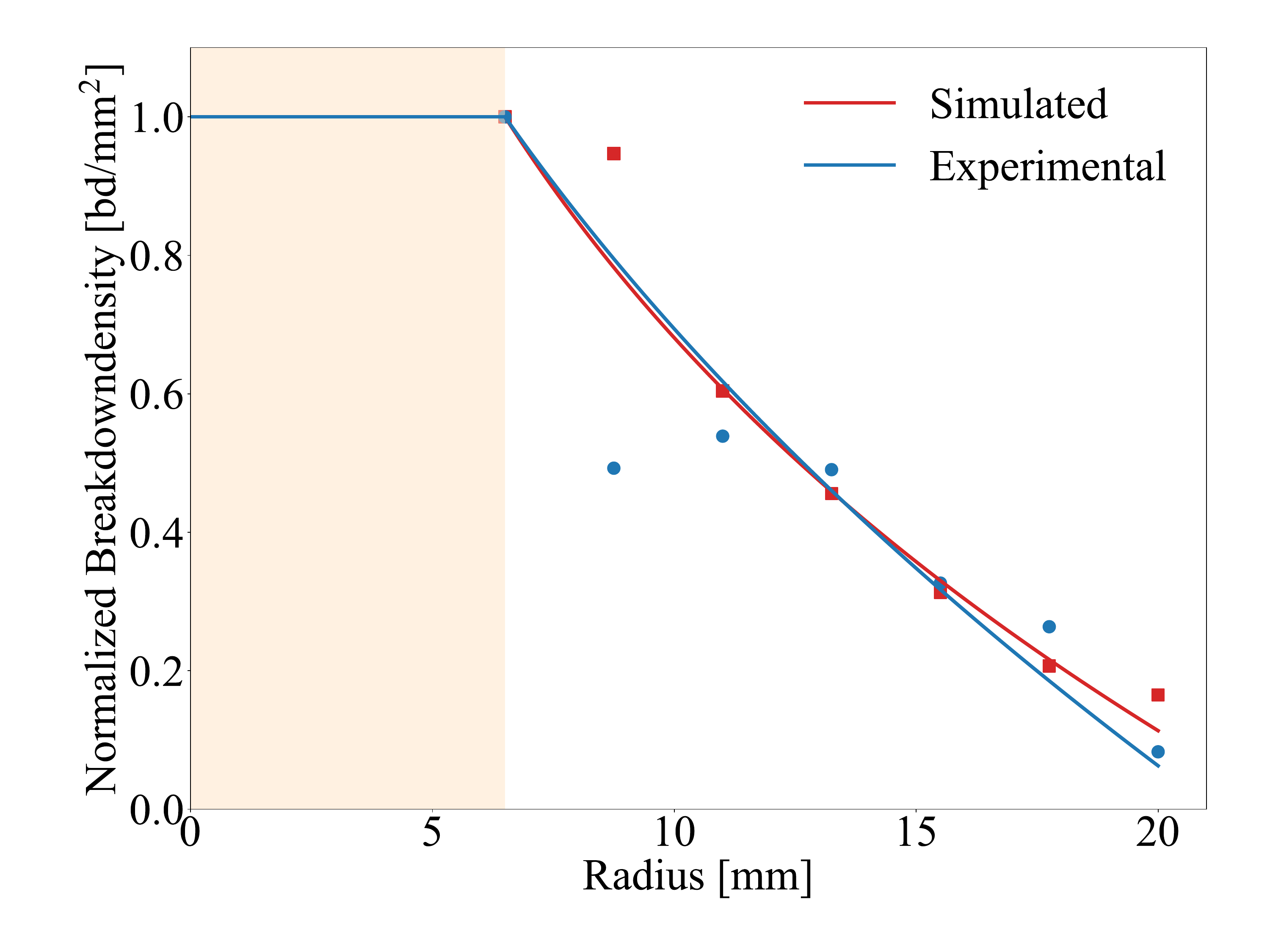}
    \caption{Polynomial fits to BDD over the radius of the electrode. The blue circles and red squares are the discrete experimental and simulated points, respectively. The blue and red lines are the fitted exponential decays. The scaling factor for the experimental data was found to be $\beta_e = 1.35$ and $\beta_s = 1.31$ for the simulated.}
    \label{fig5:CompareTosimulated}
\end{figure}

Because the surface electric field is inversely related to the radial distance as $E \propto 1/r$ when $r > 6.5$~mm and constant for $r\leq 6.5$~mm, the BDD can be found as
\begin{equation}
    BDD = 
    \begin{cases}
    const, & r\leq 6.5~mm \\
    \alpha (r-6.5~mm) ^{-\beta}, & r > 6.5~mm
     \end{cases}
\end{equation}
where $\alpha$ is a constant, removed by normalization, and $\beta$ is the scaling factor. Figure~\ref{fig5:CompareTosimulated} shows the experimental and simulated BDDs where the electrode is divided into 7 sections. The best fits of the experimental and simulated decays were found to be $\beta_e = 1.35$ and $\beta_s = 1.31$, respectively. 

Since the simulations are configured to record 1000 breakdown events before ending. In contrast, the experiments exhibit a significantly lower number of breakdown occurrences. This results in a lack of statistics for the experimental electrodes and limits the number of sections into which the electrode surface can be divided. In Figure~\ref{fig5:CompareTosimulated}, the electrode was divided into 7 sections, where the outer 6 have equal radial distance between them. No change in the results was observed if the area was divided into sections having equal surface area instead.

\section{Proposed new discussion combined with conclusion(?)}

We will now draw the main insights from the experimental data and the corresponding Monte Carlo simulations, as we have demonstrated a new approach to testing conditioning and its effects at a local scale. Figure~\ref{fig5:CompareTosimulated} clearly shows the effect of field-dependent conditioning by applying electrodes to varying electric fields, we experimentally examined the relationship between the conditioning effect of high-field exposure and breakdowns.

As the electrode's center experiences the highest electric field, it conditions more quickly than the surrounding areas. As region A becomes more resistant to vacuum breakdowns, the spatial breakdown distribution broadens, increasing the probability of breakdowns in both regions B and C. The high-field regions become dominant in the conditioning process, influencing it. This becomes evident in Fig.~\ref{fig4:Division3}, where the average final electric field between regions A and C is less than 10~MV/m, a difference of less than 10~$\%$, while the BDDs for each section differ markedly at the end of the conditioning process. The two innermost regions surpass the final BDD$_\text{C}$ at significantly lower fields, with region A surpassing at 20~MV/m and region B at 62~MV/m (indicated by black dots). 

In fact, similar observations have been reported for soft and hard copper~\cite{PhysRevAccelBeams.23.033102}. Soft copper, also known as heat-treated copper, requires a longer conditioning time than untreated copper. On the hard copper, vacuum breakdowns were concentrated in the high-field regions, leaving the lower-field regions undamaged. By contrast, on the soft copper electrode, the breakdowns were more uniformly distributed, with more breakdowns occurring in the lower-field regions. This is a direct consequence of the higher field region having more time to condition, becoming more resilient to breakdowns, and is the same result observed in this paper. In fact, current work is underway to investigate subsurface features that can help our understanding of the conditioning process. Clear dislocation activity has been observed on the surfaces of conditioned electrodes and decays spatially as one moves toward lower-field regions.

In more general terms, the agreement between Monte Carlo simulation and experiment shows that there should be a physical manifestation of $E_{state}$, either on the electrode surface or in the sub-surface material structures. The results so far do not give a clear insight into what this might be. The dislocation dynamics model, which successfully predicts the dependence of the breakdown rate on field and temperature, could serve as such a mechanism. In this case, the dislocation population in the sub-surface would evolve, for example, mobile dislocations being pinned as conditioning progresses. Such a process would occur more quickly under the increased tensile stress on the surface due to the increased applied field, as required for a physical manifestation of $E_{state}$.

\newpage
\bibliography{apssamp}

\end{document}